\documentclass{sigchi-ext}
\usepackage[T1]{fontenc}
\usepackage{textcomp}
\usepackage[scaled=.92]{helvet} 
\usepackage{graphicx} 
\usepackage{balance}  
\usepackage{booktabs} 
\usepackage{ccicons}  
\usepackage{ragged2e} 
\usepackage{amsmath}
\usepackage{setspace}

\usepackage{xspace}
\usepackage{amssymb}
\usepackage{ifthen}
\usepackage[normalem]{ulem} 
\usepackage{xcolor}

\newboolean{showedits}
\setboolean{showedits}{true} 
\ifthenelse{\boolean{showedits}}
{
	\newcommand{\del}[1]{\textcolor{red}{\sout{#1}}} 
}{
	\newcommand{\del}[1]{} 
	
}
\newboolean{showcomments}
\setboolean{showcomments}{true}
\newcommand{\id}[1]{$-$Id: scg-llncs.tex 30911 2010-02-05 10:21:47Z oscar $-$}

\ifthenelse{\boolean{showcomments}}
{\newcommand{\nbc}[3]{
 {\colorbox{#3}{\bfseries\sffamily\scriptsize\textcolor{white}{#1}}}
 {\textcolor{#3}{\sf\small$\blacktriangleright$\textit{#2}$\blacktriangleleft$}}}
 }
{\newcommand{\nbc}[3]{}
 \renewcommand{\del}[1]{} 
 }

\newcommand{\ie}{\emph{i.e.},\xspace}
\newcommand{\eg}{\emph{e.g.},\xspace}

 \usepackage{marginnote} 

\def\plaintitle{Toward Agile Situated Visualization: An Exploratory User Study} 

\def\emptyauthor{}
\def\plainkeywords{Situated Visualization; Augmented Reality; User Study;}

\title{Toward Agile Situated Visualization: An Exploratory User Study}

\numberofauthors{7}
\author{%
  \alignauthor{%
    \textbf{Leonel Merino}\\
    \affaddr{University of Stuttgart} \\
    \email{leonel.merino@visus.uni-stuttgart.de} }\alignauthor{%
    \textbf{Alexandre Bergel}\\
    \affaddr{ISCLab, Department of Computer Science,}\\
    \affaddr{University of Chile}\\
    \email{abergel@dcc.uchile.cl}\\} \vfil \alignauthor{%
    \textbf{Boris Sotomayor-G\'omez}\\
    \affaddr{Ernst Str\"{u}ngmann Institute for Neuroscience in Cooperation with Max Planck
Society}\\
    \email{boris.sotomayor@esi-frankfurt.de} }\alignauthor{%
    \textbf{Michael Sedlmair}\\
    \affaddr{University of Stuttgart} \\
    \email{michael.sedlmair@visus.uni-stuttgart.de}} \vfil \alignauthor{%
    \textbf{Xingyao Yu}\\    
    \affaddr{University of Stuttgart} \\
    \email{xingyao.yu@visus.uni-stuttgart.de} }\alignauthor{%
    \textbf{Daniel Weiskopf}\\
    \affaddr{University of Stuttgart} \\
    \email{daniel.weiskopf@visus.uni-stuttgart.de} }\vfil \alignauthor{%
    \textbf{Ronie Salgado}\\    
    \affaddr{{University of Chile} \\
    \email{roniesalg@gmail.com} } } }

\definecolor{linkColor}{RGB}{6,125,233}
\hypersetup{%
  pdftitle={\plaintitle},
  pdfauthor={\emptyauthor},
  pdfkeywords={\plainkeywords},
  bookmarksnumbered,
  pdfstartview={FitH},
  colorlinks,
  citecolor=black,
  filecolor=black,
  linkcolor=black,
  urlcolor=linkColor,
  breaklinks=true,
}

\copyrightinfo{\scriptsize 
}

\begin{document}


\maketitle

\RaggedRight{} 

\begin{abstract}
We introduce \emph{AVAR}, a prototypical implementation of an agile situated visualization (SV) toolkit targeting liveness, integration, and expressiveness. 
We report on results of an exploratory study with AVAR and seven expert users. In it, participants wore a Microsoft HoloLens device and used a Bluetooth keyboard to program a visualization script for a given dataset. 
To support our analysis, we \emph{(i)}~video recorded sessions, \emph{(ii)}~tracked users' interactions, and \emph{(iii)}~collected data of participants' impressions. 
Our prototype confirms that agile SV is feasible. That is, \emph{liveness} boosted participants' engagement when programming an SV, and so, the sessions were highly interactive and participants were willing to spend much time using our toolkit (\ie median $\geq$ 1.5 hours).
Participants used our \emph{integrated} toolkit to deal with data transformations, visual mappings, and view transformations without leaving the immersive environment. Finally, participants benefited from our \emph{expressive} toolkit and employed multiple of the available features when programming an SV. 
\end{abstract}

\ccsdesc[500]{Human-centered computing~Visualization design and evaluation methods}
\ccsdesc[300]{Human-centered computing~Mobile devices}
\keywords{\plainkeywords}

\printccsdesc

\section{Introduction}
\begin{marginfigure}[0pc]
  \begin{minipage}{0.95\marginparwidth}
    \centering
    \captionsetup{type=figure}
    \includegraphics[width=0.95\marginparwidth,height=8cm]{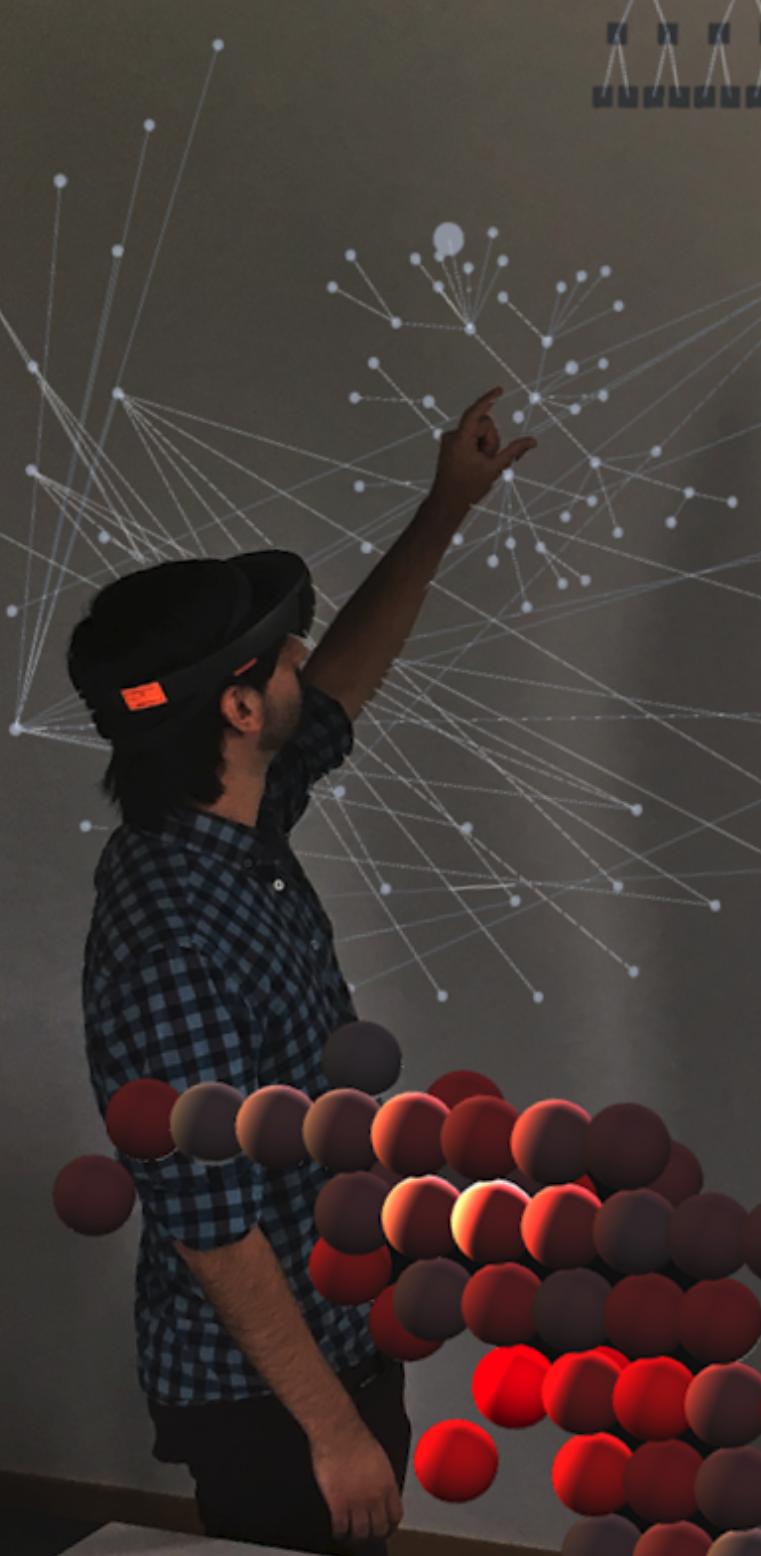}
    \caption{A user wears a Microsoft HoloLens device to interact with an SV.}~\label{fig:boris}
    ~~
    \includegraphics[width=0.8\marginparwidth]{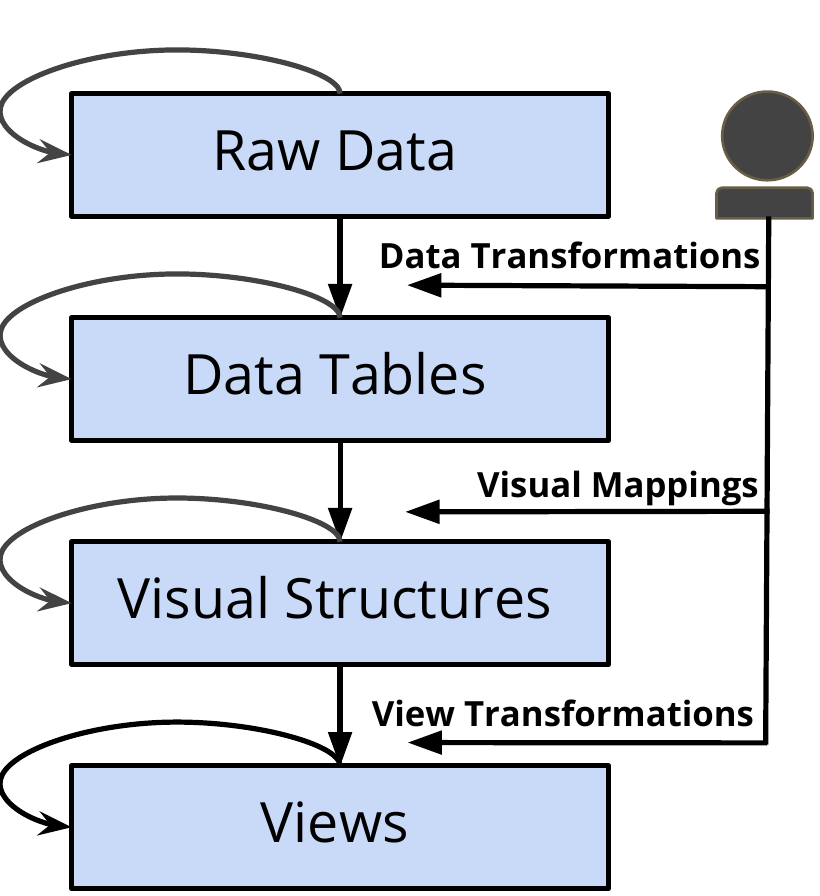}
    \caption{The 3-process reference model of visualization.}~\label{fig:model}
  \end{minipage}
\end{marginfigure}

Situated visualization (SV) promotes interactive analytical reasoning by embedding data visualizations in the physical environment through immersive augmented reality (AR)~\cite{elsayed2016situated} (see Figure~\ref{fig:boris}). Users can interact with an SV using the third spatial dimension, which stimulates cognitive aspects such as engagement, embodiment, and recall~\cite{Card97a,Marr18a}. To boost reasoning, an SV toolkit has to support users in quickly building visualizations that combine real objects 
with visual representations of data.
Yet, we observe that existing SV toolkits lack such agility, which hinders the applicability of SV in practice.
\vspace{-0.2cm}

To enable users to create and modify data visualizations situated in a real context in an agile fashion, we hypothesize that an SV toolkit must offer \emph{(i)}~\emph{live} feedback when they create or modify a visualization, \emph{(ii)}~an \emph{integrated} environment that supports all processes involved in visualization, and \emph{(iii)}~\emph{expressive} means to support increment visualization design. We implemented a prototype that supports agile SV, which we call \emph{AVAR}.
\vspace{-0.2cm}

We conducted an exploratory study with seven expert users and carefully analyzed their behavior when creating an SV. 
We observed that \emph{live} feedback boosted participants' engagement when programming an SV, and so, they were highly interactive and willing to spend long time spans (\ie median $\geq$ 1.5 hours) performing incremental modifications to visualization scripts. 
Our \emph{integrated} toolkit allowed participants to deal with the visualization as a whole 
without leaving the immersive environment. 
Finally, our \emph{expressive} toolkit allowed participants to employ multiple available features when programming an SV. 
\vspace{-0.2cm}

Our contributions are \emph{(i)} an exploratory user study and \emph{(ii)}~an open source prototype released under MIT license, thus making it fully available to practitioners and researchers\footnote{\url{https://github.com/bsotomayor92/AVAR-unity}}.

\section{Related work}
There is a lack of toolkits to guide authoring SVs that offer ready-to-use building blocks to speed up development~\cite{Marr18a}. Amongst the few existing ones, none of them focuses on agility (\ie incremental visualization construction). For instance, 
SiteLens~\cite{white2009sitelens} is a situated analytics system for supporting site visits in urban planning. In it, users can visualize an already curated dataset with a limited number of techniques. 
Munin~\cite{badam2014munin} is a middleware for ubiquitous analytics that focuses on large scale distributed visualization for collaborative environments. In it, visualizations can be displayed in mediums such as wall displays, smartphones, and tabletops, but not in immersive devices for AR (\eg Microsoft HoloLens). 
VRIA~\cite{butcher2020vria} is a Web-based framework for creating immersive analytics experiences that involves a programming language with limited capabilities of expressiveness and reflection, posing a barrier for data transformations.   
\vspace{-0.2cm}

As opposed to our SV toolkit, existing toolkits for immersive analytics such as NiwViw~\cite{yim2018niwviw}, DXR~\cite{Sica19a}, IATK~\cite{Cord19a}, and ImAxes~\cite{Cord17a} support a limited number of fixed and ready-to-use templates for visualization techniques, impairing expressiveness. These toolkits offer only partial integration, and so, they require users to perform data transformations using a desktop computer, which hinders agility. 
\vspace{-0.2cm}

Our approach differs from previous works as it targets users with programming knowledge. This fundamental difference can explain limitations of existing authoring toolkits. To the best of our knowledge, our SV approach is the first one that  
supports the three processes of programming an interactive visualization (shown in Figure~\ref{fig:model}). We consider that integrating support for all these processes in a live and expressive programming environment can lead to agile SV. 


\section{AVAR: An Agile Situated Visualization Toolkit}
Agile SV is a highly dynamic iterative process for exploring multiple facets of a dataset. In it, for instance, users can add previously filtered data to a view and analyze how it changes 
without having to leave the immersive environment (\eg removing an AR headset): script building, data exploration, and visualization exploitation happen in the AR environment. We identify three main challenges for agile SV: \emph{(C.1)} an infrastructure that supports live programming, \ie short feedback loop when evaluating a (visualization) script, \emph{(C.2)} an integrated immersive environment in which users can type and read visualizations scripts, and \emph{(C.3)} a language that is sufficiently expressive to define multiple visualization designs but simple enough to be used in immersive AR. 
\vspace{-0.2cm}

\begin{marginfigure}[-7pc]
  \begin{minipage}{0.95\marginparwidth}
    \centering
    \captionsetup{type=figure}
    \includegraphics[width=0.95\marginparwidth,height=14cm]{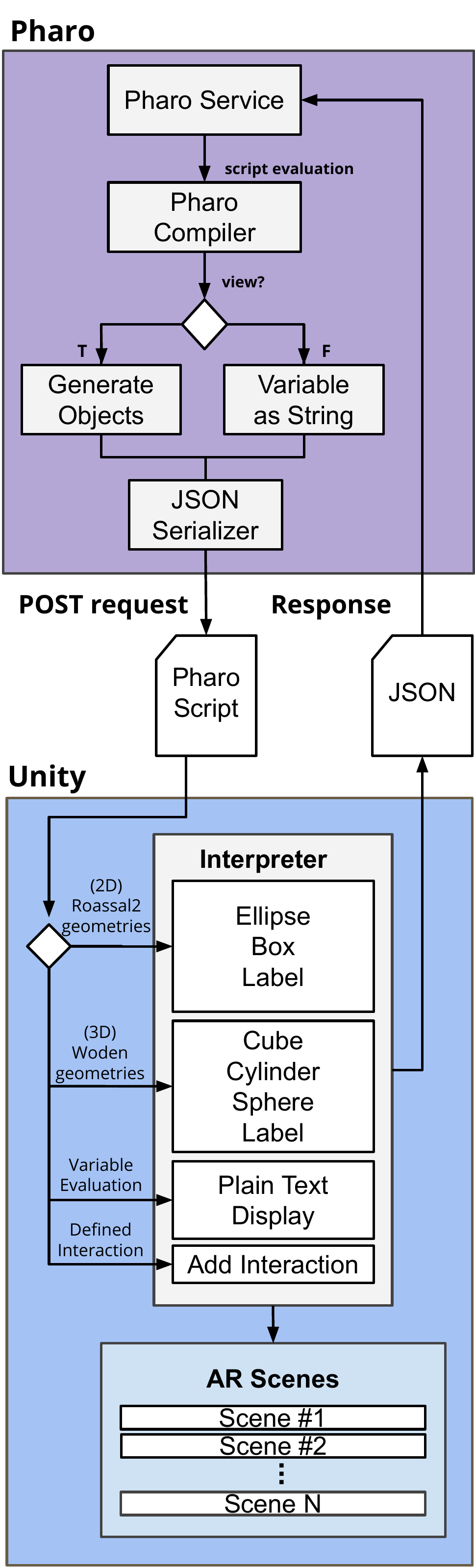}
    \caption{The distributed architecture of AVAR. 
      }~\label{fig:avar}
  \end{minipage}
\end{marginfigure}

We introduce \emph{AVAR}, our prototypical implementation for agile SV. Figure~\ref{fig:avar} presents a diagram with the software stack and the components used in the implementation of AVAR. We employ a Microsoft HoloLens and a Bluetooth keyboard as input/output devices that support user interaction, see Figure~\ref{fig:user}. 
Figure~\ref{fig:interface} shows a virtual panel that enables users to type visualization scripts, load visualization examples, and receive error notifications.  
\vspace{-0.2cm}

In the design of AVAR, we maximize reusing existing tools. As a consequence, the implementation phase mainly consists of integrating these third-party tools in the immersive AR environment. We integrate a fully operational programming language into an immersive environment. To this end, we adopt Pharo\footnote{\url{http://pharo.org/}, accessed 15.12.2019}, a modern implementation of Smalltalk. Pharo is a dynamically typed message passing language that has an expressive syntax that allows users to perform complex operations by typing short scripts~\cite{Berg16c}. Pharo is interpreted, that is, users do not have to wait for a script to compile, but they can evaluate scripts in a live environment. Pharo is highly reflective, which eases integration to external environments. Users can define data visualizations using Roassal2\footnote{\url{https://github.com/ObjectProfile/Roassal2}, accessed 15.12.2019} and Woden\footnote{\url{https://github.com/ronsaldo/woden}, accessed 15.12.2019}, 2D and 3D data visualization engines, respectively. These engines implement multiple 2D and 3D visualization techniques that are shipped out-of-the-box such as parallel coordinates, treemaps, node-link diagrams, and space-time cube matrices. Finally, we implement a thin application in Unity 3D\footnote{\url{https://unity3d.com/}, accessed 15.12.2019} that communicates with Pharo as a backend, handles user interaction, and renders a graphical user interface in AR. 
Our toolkit supports the following 3-step process:
\begin{enumerate}\compresslist%
    \item \emph{Data Transformations}. 
    To build a data visualization, users first apply several transformations to a given raw dataset to create data tables, \eg filtering, formatting, normalizing. As these transformations are available in Pharo, users have access to multiple functionalities for data transformations. To the best of our our knowledge, this process is not fully supported by existing SV toolkits.
    \item \emph{Visual Mappings}. 
    Next, users choose visual mappings to apply to data tables toward creating visual structures. To this end, we rely on multiple existing "builders" in Roassal2 and Woden, which are domain-specific languages (DSLs) that support the rapid construction of particular interactive visualizations.
    \item \emph{View Transformations}. 
    Finally, a view is rendered in immersive AR to which users apply various view transformations (programmatically as well as using natural user interfaces). For instance, through hand gestures users can rotate a view to obtain a different perspective. Users can also combine a hand gesture with walking to relocate a view to a different place. 
\end{enumerate}

\section{Exploratory User Study}
We adopted a template previously introduced~\cite{Meri18a} to describe the scope of our study: 
\vspace{-1em}

\framebox{
	\parbox[t][1.6cm]{0.93\linewidth}{
	\addvspace{0.01cm}
       We examine the usage of the AVAR toolkit for authoring \emph{situated visualizations} displayed in a \emph{Microsoft HoloLens} device in the context of \emph{an exploratory analysis} from the point of view of \emph{expert users}.
	} 

}\\

\begin{marginfigure}[0pc]
  \begin{minipage}{0.95\marginparwidth}
    \centering
    \captionsetup{type=figure}
    \includegraphics[width=0.93\marginparwidth]{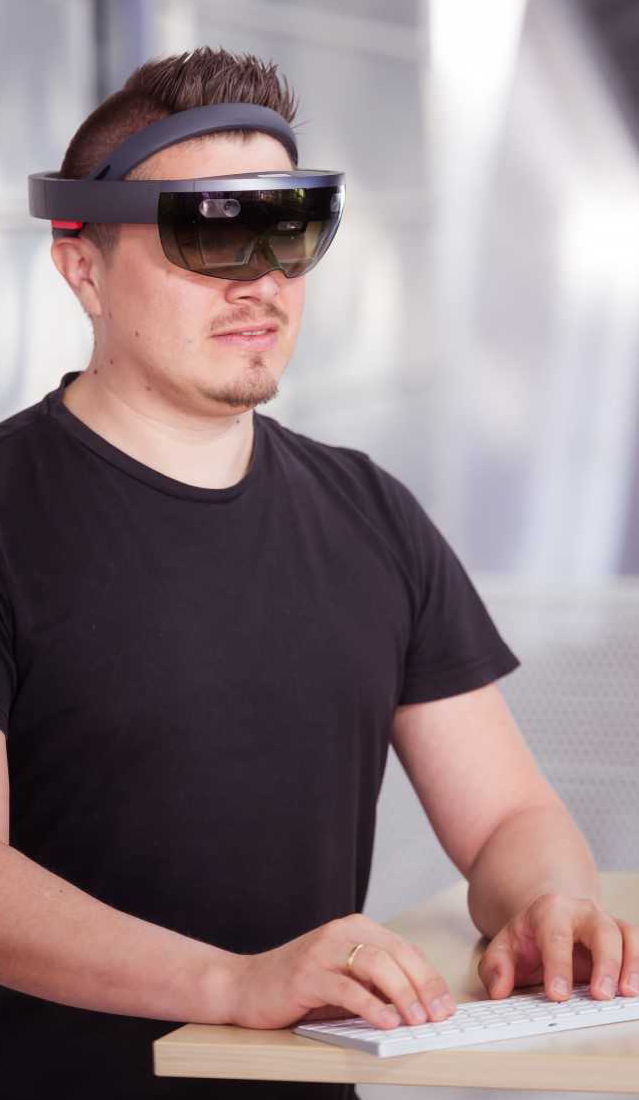}
    \caption{In the study, participants wore a Microsoft HoloLens device and used an Apple Magic Bluetooth keyboard. 
      }~\label{fig:user}
    \includegraphics[width=0.93\marginparwidth]{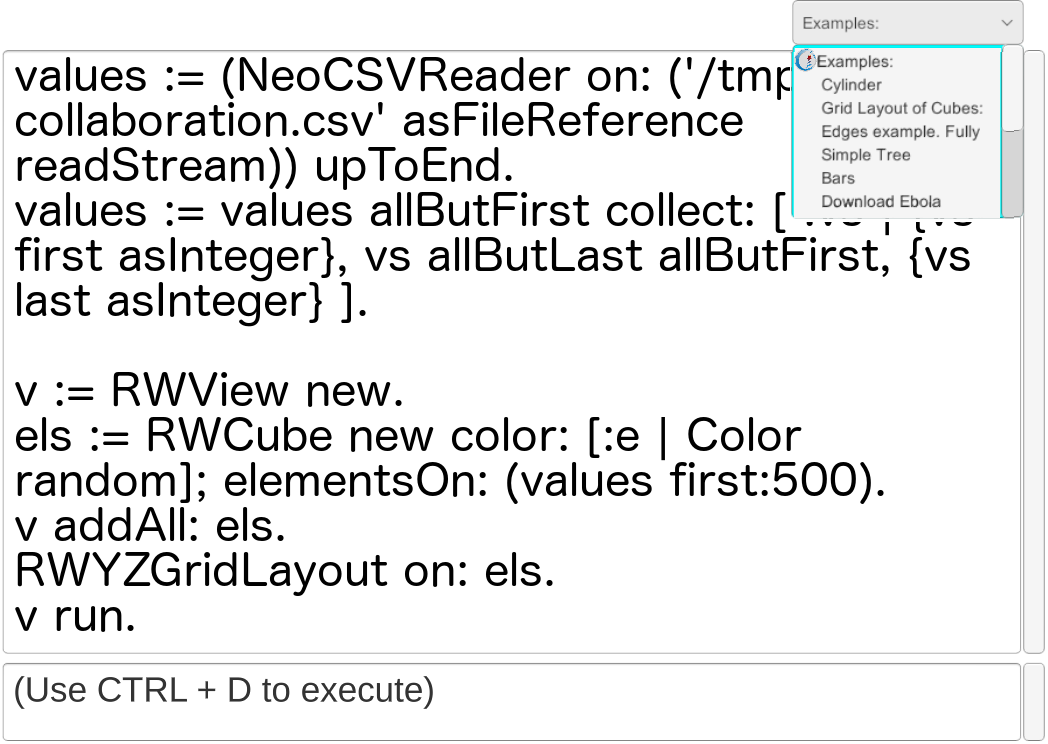}
    \caption{The graphical user interface in AR of AVAR: (center) code editor, (upper-right corner) examples browser, and (bottom) console panel.
      }~\label{fig:interface}
  \end{minipage}
\end{marginfigure}
\vspace{0.3em}

\noindent\textbf{{Pilot}}
We used a pilot study with two participants to fine-tune the study factors: task and datasets that could be understood quickly, and the inclusion of more examples to demonstrate the capabilities of visualization engines.  
\vspace{-0.2cm}

\noindent\textbf{Participants}
As our prototype uses a Smalltalk scripting language, we decided to conduct our study at ESUG'19\footnote{European Smalltalk Users Group, accessed December 15, 2019, \url{https://esug.github.io/2019-Conference/conf2019.html}}. We sent an open invitation through the conference mailing list. In the end, we scheduled sessions with seven participants, who were not paid and freely opted to participate in the study. All participants were male. Their median age was 31 \raisebox{.2ex}{$\scriptstyle\pm$} 8.5 years, and they had a considerable experience using Smalltalk (\ie experience $\geq$ 6 years). We also asked for their familiarity with the technologies involved in the study. In summary, participants declared to \emph{(i)} frequently build data visualizations, \emph{(ii)} know little of the API of the 2D visualization engine, \emph{(iii)} do not know details of the API of the 3D visualization engine, and \emph{(iv)} have used a device like the Microsoft HoloLens no more than once.
\vspace{-0.2cm}

\noindent\textbf{Dataset}
We selected two datasets used in previous studies~\cite{Sica19a,Cord17a}. One dataset contains co-authorship information along a period of time, which we considered adequate to minimize the complexity of Task 1 (\eg it has 209 items and 4 properties). In Task 2, participants used a second dataset that contains  6,497 samples of wine (1,600 red and 4,897 white) described by 12 data attributes. Both datasets are publicly available\footnote{\url{https://github.com/ronellsicat/DxR/blob/master/Assets/StreamingAssets/DxRData/collaboration.csv}, accessed 15.12.2019}\textsuperscript{,}\footnote{\url{http://www3.dsi.uminho.pt/pcortez/wine/winequality.zip}, accessed 06.01.2020}.
\vspace{-0.2cm}

\begin{marginfigure}[-3pc]
  \begin{minipage}{0.95\marginparwidth}
  \vspace*{-0.9cm}
    \centering
    \includegraphics[width=0.93\marginparwidth,height=2.7cm]{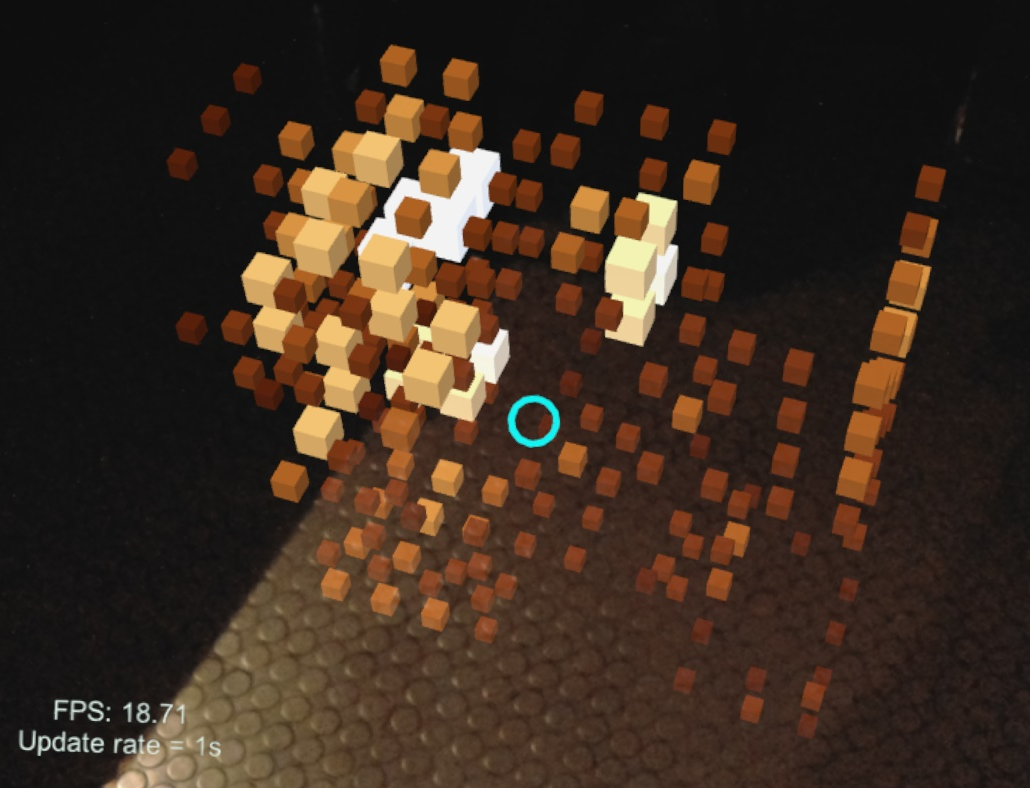}
    \includegraphics[width=0.93\marginparwidth,height=2.7cm]{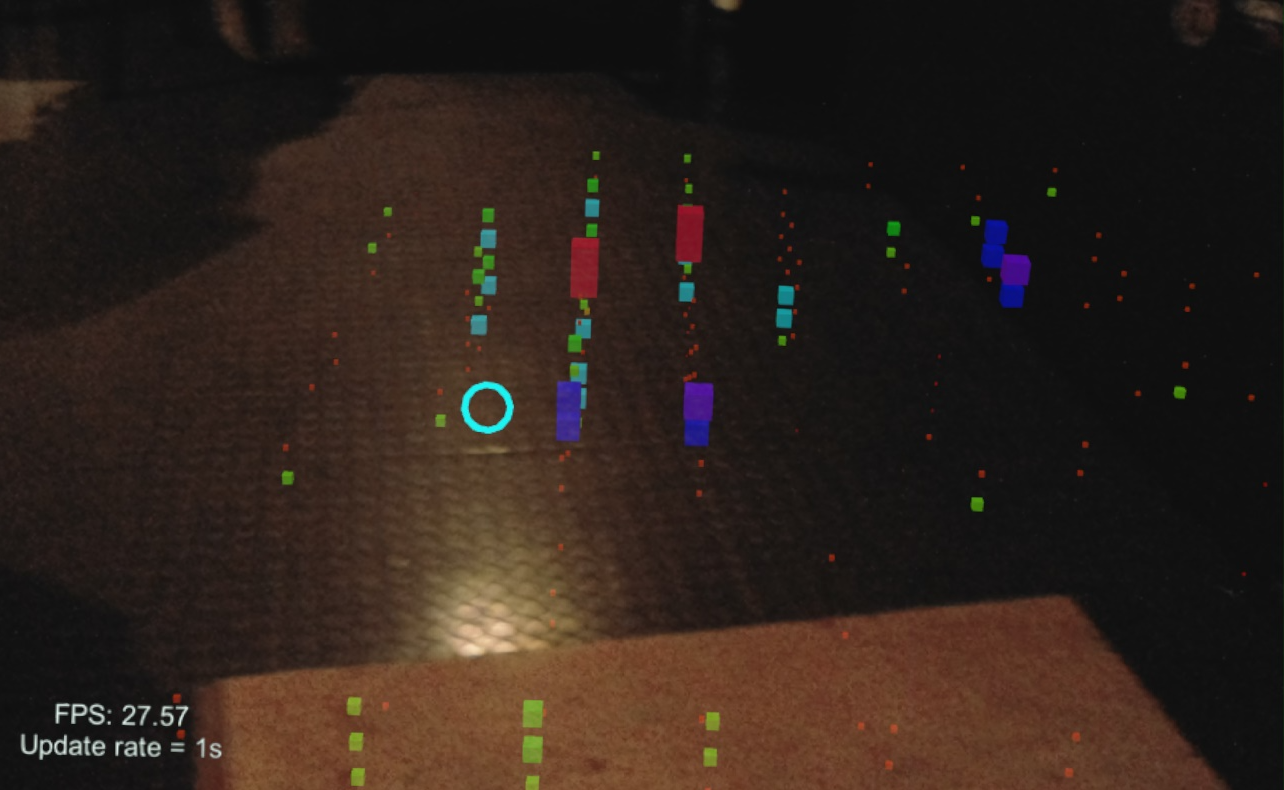}
    \includegraphics[width=0.93\marginparwidth,height=2.7cm]{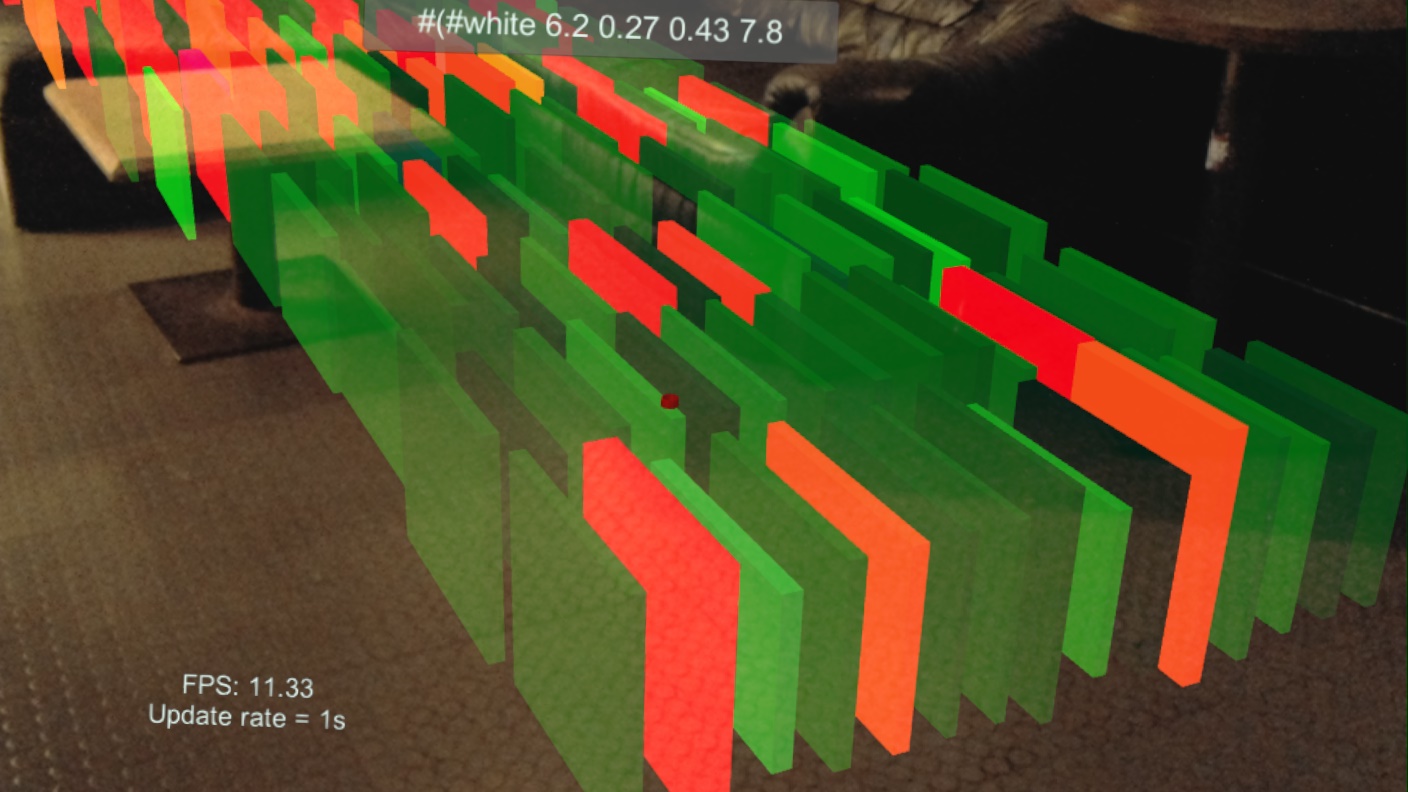}
    \caption{Visualizations by participants in the user study. 
      }~\label{fig:vis}
  \end{minipage}
\end{marginfigure}

\noindent\textbf{Tasks}
\emph{Task 1. }We asked participants to build a space-time cube visualization (results shown in Figures~\ref{fig:vis}). Each cube represents the relation between two co-authors. The $X$ and $Z$ axes (co-planar to the room's floor) represent the list of authors, and the $Y$ axis represents time. Time is overloaded in the color of the cubes, which use a color ramp from blue to yellow. To clarify the given task, we handed to participants a printout of the expected resulting visualization. \emph{Task 2.} As a second, and optional task, we asked participants to analyze main differences between white and red wine based on the given dataset. To this end, participants were encouraged to use the multiple features available in the visualization engines.    
\vspace{-0.2cm}

\noindent\textbf{Apparatus}
Participants wore a Microsoft HoloLens 1 headset. 
The headset was complemented with an Apple Magic Bluetooth keyboard (see Figure~\ref{fig:user}).
Participants used the keyboard to interact with the three main sections of the graphical user interface: \emph{(i)}~code editor, \emph{(ii)}~examples browser, and \emph{(iii)}~console panel (as shown in Figure~\ref{fig:interface}). Participants could scroll through the code either  using the keyboard or a hand gesture. Participants could interact with visualizations in three ways by \emph{(1)} hovering over an element using head movements to obtain contextual information, \emph{(2)} rotating a visualization using an airtap and hold combined with an horizontal hand gesture, and \emph{(3)} translating the visualization to a new location by using an airtap hand gesture and body and head movements. 
\vspace{-0.2cm}

\noindent\textbf{Procedure}
The sessions with each participant were conducted in a quiet room. The room had enough space for the participants to move around. The room had a high table for stand-up coding and a normal table for participants who preferred to sit on a chair. At the beginning of the session, participants were asked to read and sign a consent form that informed them of the characteristics of the study. Next, the experimenter read an introduction to explain the details of the phases in the study. 
We encouraged participants to share their thoughts using a think-aloud protocol.
Participants were free to stop the session at any time. Once participants finished a task, we asked them about the perceived difficulty of the task. To examine the (lack of) comfort experienced by participants, we asked them to fill in a Simulator Sickness questionnaire~\cite{Kenn03a}. To assess the perceived usability of our system, participants were asked to fill in a System Usability Scale questionnaire~\cite{Broo95a}. 
\vspace{-0.2cm}
\begin{margintable}[1pc]
 \begin{minipage}{\marginparwidth}
    \centering
    \vspace{-2em}
    \footnotesize
    \setlength\tabcolsep{3pt}
\renewcommand{\arraystretch}{0.8}
    \begin{tabular}{ll}
    {\footnotesize \textbf{Simulator Sickness}}  & {\footnotesize \textbf{Rating}} \\
      \toprule
        General discomfort & Moderate\\
        Fatigue & Slight\\
        Headache & None\\
        Eye strain & Slight\\
        Difficulty focusing & Slight\\
        Increased salivation & None\\
        Sweating & None\\
        Nausea & None\\
        Difficulty concentrating & None\\
        Fullness of the head & Slight\\
        Blurred vision & Slight\\
        Dizzy (eyes open) & None\\
        Dizzy (eyes closed) & None\\
        Vertigo (Giddiness) & None\\
        Stomach awareness & None\\
        Burping & None\\
      \bottomrule
    \end{tabular}
    \caption{Median ratings using the Simulator Sickness questionnaire.}~\label{tab:table1}
  \end{minipage}
\end{margintable}
\noindent\textbf{Data Collection}
We \emph{(i)} video recorded the sessions (\ie 13 hours and 30 minutes), \emph{(ii)} tracked events of participants' interactions with the graphical user interface and with the environment (\ie 1029 interaction events), and \emph{(iii)} collected filled-in questionnaires (\ie 28 pages in total).

\section{Results}
\begin{marginfigure}[0pc]
  \begin{minipage}{0.95\marginparwidth}
    \centering
    \includegraphics[width=0.93\marginparwidth]{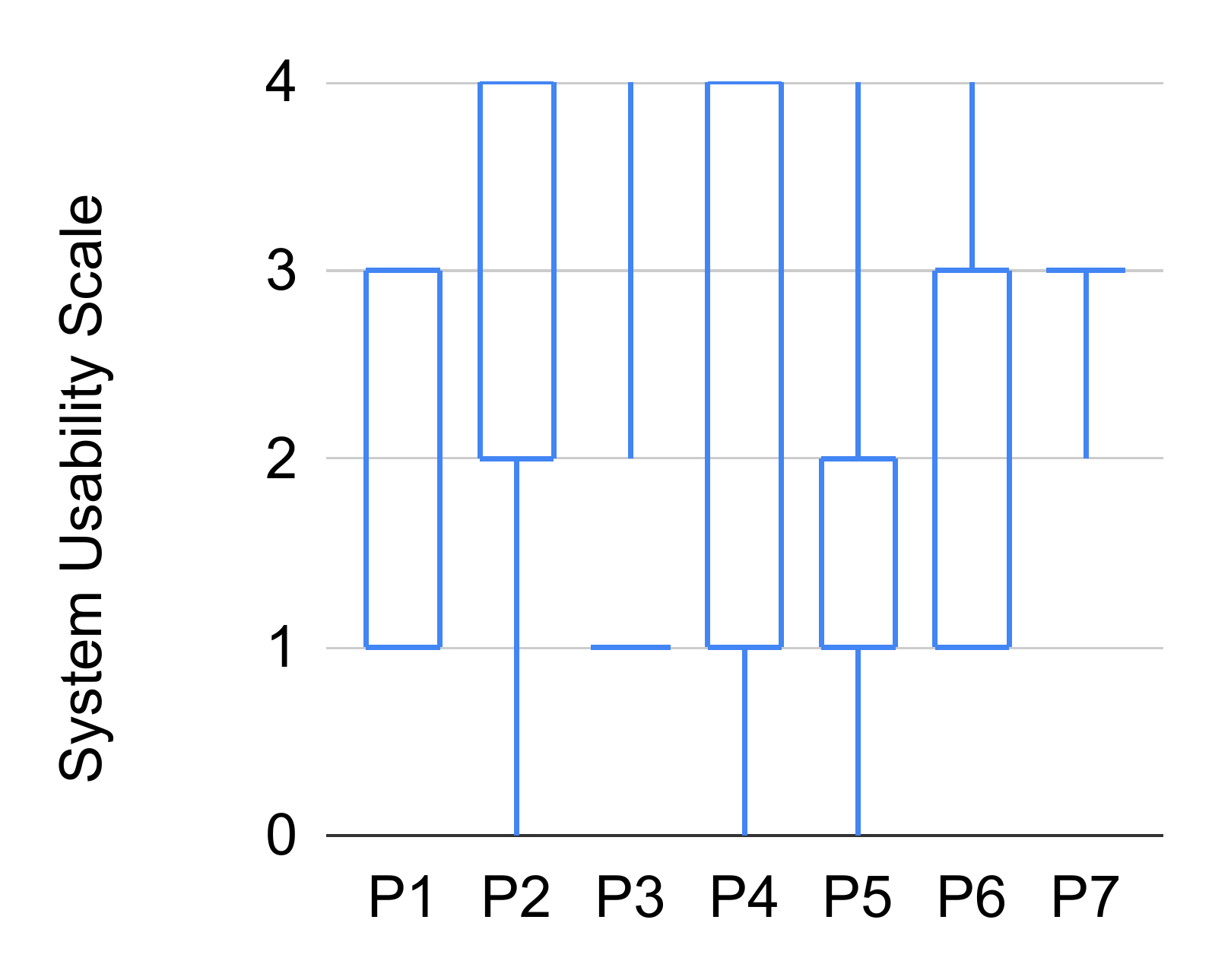}
    \caption{Ratings by participants using the System Usability Scale questionnaire. 
      }~\label{fig:sus}
  \end{minipage}
\end{marginfigure}

A set of charts that summarize the results of the study is presented in Figure~\ref{fig:results}. Due to the limited space, we opted to present only the results of \emph{Task 1}, even though 5 participants also solved \emph{Task 2}. Charts are sorted by time (\eg participant \textcircled{\small{1}}, who had the longest session, is presented at the top). A horizontal black bar, at the middle of each chart, encodes the length of a session. Such bars split charts into two sections. In the upper section, gray circles are vertically arranged to indicate which area in the graphical interface has the focus of a participant. A green circle indicates a script that is successfully executed, otherwise, the circle is red. An additional horizontal bar encodes, using three colors, which visualization process participants are addressing. The lower section of a chart supports a temporal analysis of visualization scripts: Vertical bars (in green) show additions and (in red) deletions of code. Additional circles depict the total size of a script at certain points in time. Circles are connected with lines to indicate the evolution of the script size. The median values of the ratings of participants using the Simulator Sickness questionnaire are presented in Table~\ref{tab:table1} and results of  the System Usability Scale (SUS) questionnaire are presented in Figure~\ref{fig:sus}. The median SUS score by participants was 58, with a maximum of 70 and a minimum of 53. The median rate at which participants interacted with our system was 1.5 \raisebox{.2ex}{$\scriptstyle\pm$} 0.7 events per minute.

\section{Discussion}
AVAR confirms the feasibility of an agile SV toolkit based on liveness, integration, and expressiveness. All participants were able to solve the first task, even though they experienced moderate discomfort wearing the headset for a long period of time. Participants agreed with the observation that "most people would learn to use the system quickly" as they did not need to learn lots of things to get going with the system. However, they considered that AVAR requires improvements to be easy-to-use.   
\vspace{-0.2cm}

\begin{figure}[t]
	\centering
	\captionsetup{type=figure}
	\includegraphics[width=\linewidth, height=0.9\textheight]{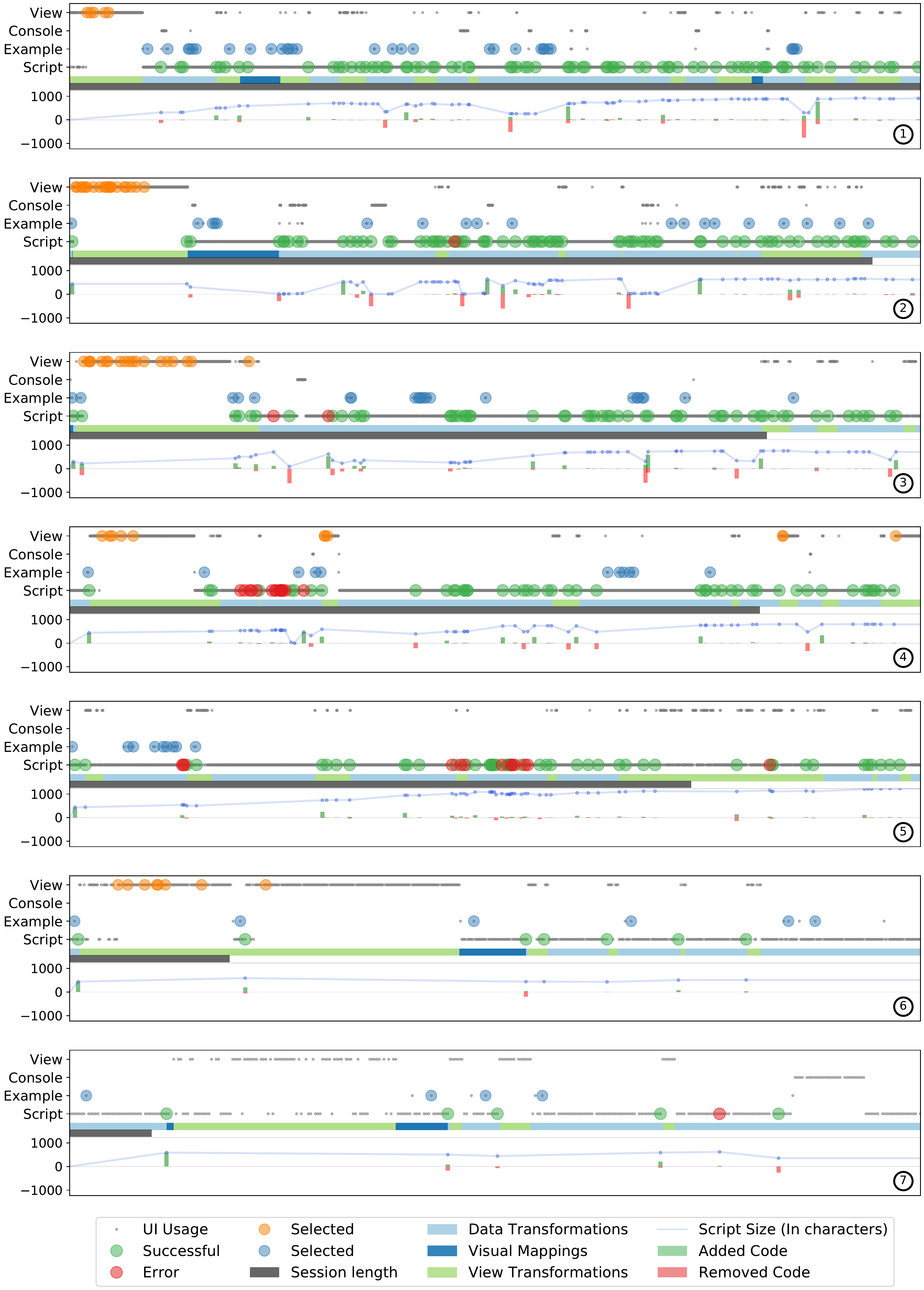}
	\caption{Charts that present the interactions of the seven participants who used our SV toolkit.}
	\label{fig:results}
\end{figure}

\textbf{Liveness.} More experienced users (\eg \textcircled{\small{6}} and \textcircled{\small{7}}) solved the Task 1 with fewer interactions in a shorter time than less experienced users (\eg \textcircled{\small{1}} and \textcircled{\small{2}}). We also observe that our live environment boosted participants engagement, as five out of seven participants were willing to solve the second (optional) task. All sessions were highly interactive. In agile SV, users rely on liveness to obtain feedback of script executions, which indeed, were uniformly distributed in time (see Figure~\ref{fig:results}).
\vspace{-0.2cm}

\textbf{Integration.} Participants engaged in long experimental sessions that lasted a median of 106 minutes\raisebox{.2ex}{$\scriptstyle\pm$}27.5 (with a maximum of 148 minutes and a minimum of 82 minutes). In them, participants were able to deal with all the processes required to solve the tasks without leaving the immersive environment. In agile SV, users require to address visualization processes in a continuous loop (and not sequentially) as show in Figure~\ref{fig:results}.  
\vspace{-0.2cm}
\setlength\marginparsep{10cm} 

\textbf{Expressiveness.} Our included expressive language enabled participants to use various features that they found amongst the visualization engines available. For instance, participants \textcircled{\small{1}} and \textcircled{\small{3}} used \texttt{RTTabTable} to manipulate data tables, participants \textcircled{\small{2}} and  \textcircled{\small{7}} used \texttt{RWElement} for handling 3D elements, participant \textcircled{\small{2}} used \texttt{RWAlign} to layout elements in a 3D space, participant \textcircled{\small{3}} used \texttt{RWCylinder} to produce cylinders as visual elements, and participant \textcircled{\small{7}} used \texttt{RTScale} to scale elements and maximize the use of the available space in the room. Other features used by all participants were \texttt{RWView} to specify views, \texttt{RWXZGridLayout} to layout elements as a 3D grid, and \texttt{RWCube} to define cube shapes for elements. In agile SV, users depend on having multiple features available to express SV designs in an iterative fashion. 

 \setlength\marginparsep{10cm} 


 \marginpar{%
  \fbox{%
    \begin{minipage}{0.93\marginparwidth}
      \textbf{Acknowledgments} \\
      \vspace{0.1pc} 
      \footnotesize
Merino, Sedlmair, Sotomayor-G\'{o}mez, and Weiskopf  acknowledge funding by the Deutsche Forschungsgemeinschaft (DFG, German Research Foundation) -- Project-ID 251654672 -- TRR 161. Bergel thanks LAM Research for its financial support. Sotomayor-G\'omez thanks Universidad Austral de Chile and Universidad de Chile for their financial support.
    \end{minipage}} }
\section{Conclusion}
We introduce \emph{AVAR}, an agile SV toolkit based on liveness, integration, and expressiveness. We report on an exploratory study with seven expert users who were asked to program an SV script. We analyzed how our design choices impacted participants' behavior. We found that live feedback boosted the engagement of the participants, who worked highly interactively and were willing to spend much time using our toolkit. Participants were able to deal with visualization as a whole without leaving the immersive environment. Finally, we observed that participants employed multiple of the available features when programming an SV. In the future, we plan to improve our design, investigate other means for interaction, and conduct further evaluations.


%
%
%
%
%
\balance{}

\balance{}

\bibliographystyle{SIGCHI-Reference-Format}
\bibliography{avar}

\end{document}